\documentclass[a4paper, 10pt, conference]{IEEEtran}
\IEEEoverridecommandlockouts

\usepackage{cite}
\usepackage{amsmath,amssymb,amsfonts}

\usepackage{algorithm}
\usepackage{algpseudocode}

\usepackage[bottom,hang,flushmargin]{footmisc}

\newcommand{\Input}{\State\textbf{Input:} }
\newcommand{\Output}{\State\textbf{Output:} }

\usepackage{diagbox}
\usepackage{booktabs}
\usepackage{colortbl}
\usepackage{multirow}
\usepackage{tabularx}
\usepackage{soul}
\usepackage{wrapfig}
\usepackage[caption=false]{subfig}
\usepackage[font=footnotesize,labelfont=bf]{caption}
\usepackage{dblfloatfix}

\usepackage{enumitem}
\usepackage{flushend}

\usepackage{layout}

\usepackage{pifont}      

\newcommand{\cmark}{\ding{51}}
\newcommand{\xmark}{\ding{55}}
\newcommand{\pmark}{\raisebox{0.2ex}{\scriptsize$\sim$}} 

\newcommand{\rone}{\blacksquare\,\square\,\square\,\square\,\square}
\newcommand{\rtwo}{\blacksquare\,\blacksquare\,\square\,\square\,\square}
\newcommand{\rthree}{\blacksquare\,\blacksquare\,\blacksquare\,\square\,\square}
\newcommand{\rfour}{\blacksquare\,\blacksquare\,\blacksquare\,\blacksquare\,\square}
\newcommand{\rfive}{\blacksquare\,\blacksquare\,\blacksquare\,\blacksquare\,\blacksquare}

\usepackage{bm}
\makeatletter
\AtBeginDocument{\DeclareMathVersion{bold}
\SetSymbolFont{operators}{bold}{T1}{times}{b}{n}

\SetMathAlphabet{\mathrm}{bold}{T1}{times}{b}{n}
\SetMathAlphabet{\mathit}{bold}{T1}{times}{b}{it}
\SetMathAlphabet{\mathbf}{bold}{T1}{times}{b}{n}
\SetMathAlphabet{\mathtt}{bold}{OT1}{pcr}{b}{n}
\SetSymbolFont{symbols}{bold}{OMS}{cmsy}{b}{n}
\renewcommand\boldmath{\@nomath\boldmath\mathversion{bold}}}
\makeatother

\usepackage{graphicx}
\usepackage{textcomp}
\usepackage{xcolor}
\usepackage{flushend}
\usepackage[a4paper, total={184mm,239mm}]{geometry}
\def\BibTeX{{\rm B\kern-.05em{\sc i\kern-.025em b}\kern-.08em
    T\kern-.1667em\lower.7ex\hbox{E}\kern-.125emX}}
\begin{document}

\title{\emph{NuRedact}: Non-Uniform eFPGA Architecture for Low-Overhead and Secure IP Redaction}

\author{\IEEEauthorblockN{Voktho Das, Kimia Azar, Hadi Kamali}
\IEEEauthorblockA{\textit{Department of Electrical and Computer Engineering (ECE), University of Central Florida, Orlando, FL 32816, USA} \\
\{voktho.das , azar, kamali\}@ucf.edu}
}

\maketitle

\begin{abstract}

While logic locking has been extensively studied as a countermeasure against integrated circuit (IC) supply chain threats, recent research has shifted toward reconfigurable-based redaction techniques, e.g., LUT- and eFPGA-based schemes. While these approaches raise the bar against attacks, they incur substantial overhead, much of which arises not from genuine functional reconfigurability need, but from artificial complexity intended solely to frustrate reverse engineering (RE). As a result, fabrics are often underutilized, and security is achieved at disproportionate cost. This paper introduces \emph{NuRedact}, the first full-custom eFPGA redaction framework that embraces architectural non-uniformity to balance security and efficiency. Built as an extension of the widely adopted OpenFPGA infrastructure, \emph{NuRedact} introduces a three-stage methodology: (i) custom fabric generation with pin-mapping irregularity, (ii) VPR-level modifications to enable non-uniform placement guided by an automated Python-based optimizer, and (iii) redaction-aware reconfiguration and mapping of target IP modules. Experimental results show up to 9× area reduction compared to conventional uniform fabrics, achieving competitive efficiency with LUT-based and even transistor-level redaction techniques while retaining strong resilience. From a security perspective, \emph{NuRedact} fabrics are evaluated against state-of-the-art attack models, including SAT-based, cyclic, and sequential variants, and show enhanced resilience while maintaining practical design overheads.

\end{abstract}


\section{Introduction}

Outsourcing critical stages of IC development to global parties has unlocked scale and cost advantages, but it has also exposed designs to IP piracy, RE, and unauthorized overproduction \cite{rostami2014primer}. Logic locking emerged as a systematized countermeasure \cite{kamali2022advances}, yet attacks on logic locking (i.e., de-obfuscation attacks), both logical and physical \cite{subramanyan2015evaluating, azar2019smt, barenghi2012fault, kocher1999differential}, have eroded the practical security of many locking breeds under realistic threat models, motivating a reassessment of protection primitives \cite{kamali2022advances}. Over the last few years, a natural pivot has been shift towards reconfigurable-based \emph{redaction}, which isolates security-critical logic cones and IPs and replaces them with reconfigurable substrate \cite{kamali2018lut, kamali2019full, collini2022reconfigurable, kamali2023shell, abideen2024overview, das2025soar}, typically LUT or embedded FPGA (eFPGAs) fabrics, so that functionality is convoluted and only instantiated with a protected configuration (bitstream). 

Early redaction flows have automated module selection and clustering for either LUT-based \cite{guo2023evolute} and eFPGA-based \cite{bhandari2021exploring} redaction. However, they have fixed the target fabric architecture/placement (particularly in eFPGA-based studies), leaving significant PPA slack \cite{bhandari2021exploring, tomajoli2022alice,  bhandari2023not}. The last two years have seen a sharper focus on fabric structure and granularity \cite{bhandari2023not, kamali2023shell, abideen2024overview, dasgupta2025hipr, collini2025arianna, fowler2025trap}. HIPR \cite{dasgupta2025hipr} inserts fine-grain configurable blocks (including custom LUTs, sequential blocks, and in interconnects) for Boolean logic, sequential logic, interconnect randomization, respectively, then compacts the configuration to curb PPA/bitstream cost while maintaining resistance to functional and structural attacks. ARIANNA \cite{collini2025arianna} closes the loop between what to redact and how to build the fabric, by identifying a secure subset of eFPGA parameters up front and tailors bespoke fabrics per module-cluster. TRAP \cite{fowler2025trap} moves below the gate level by a transistor-programmable fabric whose switch-level semantics fundamentally complicate attacks (e.g., the SAT attack \cite{subramanyan2015evaluating, el2019sat, shamsi2017appsat, shen2017double, zhou2017cycsat, shamsi2019icysat}). 

While these approaches mitigate overhead, they still rely on uniform attributes that provision routing capacity, LUT structures, and tile regularity for generality rather than redaction specificity. As a result, when different logic cones are redacted, substantial configuration, logic, and routing resources remain underutilized \cite{dasgupta2025hipr, collini2025arianna, fowler2025trap}. Moreover, bitstream storage and configuration flip-flops incur nontrivial silicon area and timing costs. Such customizations may also leak structural cues, enabling oracle-less or structure-guided attacks.

Among these approaches, despite overhead concerns, eFPGAs remain attractive for integration and tooling, from open-source IP generators (e.g., OpenFPGA \cite{tang2019openfpga}), to commercial IP (Achronix Speedcore, Flex Logix EFLX, QuickLogic), being integrated in SoCs, with the value proposition of post-silicon patchability \cite{tang2019openfpga}. While eFPGAs are deployed to support future reconfiguration, in redaction (used for IP protection), these substrates are used to instantiate complexity (a large functional/search space) and to decouple revealed structure from protected behavior. This means the \emph{fabric’s non-uniformity can be co-designed with the target IP}: redaction needs \textbf{programmability as a means} (security), not as a \textbf{product feature}. 

Motivated by this fact, we propose \emph{NuRedact}, an automated framework for non-uniform eFPGA redaction, realizing the need for programmability as a means of protection at the lowest overhead, while retaining security and maximizing integration and tooling. \emph{NuRedact} replaces IP sub-components with irregular, design-specific fabrics generated atop OpenFPGA, modifies VPR to legalize non-uniform placement, and includes a redaction-aware mapper that exploits irregular pin-mapping and tailored resources. Compared to HIPR and TRAP, \emph{NuRedact} preserves mainstream RTL-to-OpenFPGA automation and industry tool compatibility at comparable overhead while cutting area by up to 9x vs. baseline eFPGA-based redaction. The contributions of \emph{NuRedact} are as follows:

\begin{table*}[t]
\centering
\footnotesize
\setlength{\tabcolsep}{4pt}
\caption{Top-level Comparison of Redaction Approaches. ($\rone$~=~low, $\rfive$~=~high). \cmark\ = present, \xmark\ = absent, \pmark\ = partial/indirect).}
\label{tab:background-comparison}
\begin{tabular}{@{}l l c c c c c l@{}}
\toprule
\textbf{Technique} & \textbf{Granularity} & \textbf{Overhead} & \textbf{Bitstream} & \textbf{Interconnect} & \textbf{Custom} & \textbf{Tooling} & \textbf{Highlight} \\
& & & & \textbf{obfuscation} & \textbf{fabric} & \textbf{\& integration} & \\
\midrule
LUT-Lock \cite{kamali2018lut} & LUT obfuscation & $\rthree$ & $\rthree$ & \xmark & \xmark & \cmark & SAT-hardening \\
\cmidrule(r){1-1}\cmidrule(r){2-2}\cmidrule(r){3-3}\cmidrule(r){4-4}\cmidrule(r){5-5}\cmidrule(r){6-6}\cmidrule(r){7-7}\cmidrule(r){8-8} 
Full-Lock \cite{kamali2019full} & Configurable Routing + LUT & $\rthree$ & $\rfour$ & \cmark & \xmark & \xmark & SAT-hardening + Proof\\
\cmidrule(r){1-1}\cmidrule(r){2-2}\cmidrule(r){3-3}\cmidrule(r){4-4}\cmidrule(r){5-5}\cmidrule(r){6-6}\cmidrule(r){7-7}\cmidrule(r){8-8} 
ALICE \cite{tomajoli2022alice} & eFPGA (fixed fabric) & $\rfive$ & $\rfour$ & \xmark & \xmark & \cmark & End-to-end Flow \\
\cmidrule(r){1-1}\cmidrule(r){2-2}\cmidrule(r){3-3}\cmidrule(r){4-4}\cmidrule(r){5-5}\cmidrule(r){6-6}\cmidrule(r){7-7}\cmidrule(r){8-8} 
SheLL \cite{kamali2023shell} & eFPGA (tile shrinking) & $\rthree$ & $\rthree$ & \pmark & \cmark & \cmark & $\downarrow\sim$55\% vs. ALICE \cite{tomajoli2022alice} \\
\cmidrule(r){1-1}\cmidrule(r){2-2}\cmidrule(r){3-3}\cmidrule(r){4-4}\cmidrule(r){5-5}\cmidrule(r){6-6}\cmidrule(r){7-7}\cmidrule(r){8-8} 
HIPR \cite{dasgupta2025hipr} & LUT + seq + interconnect & $\rtwo$ & $\rtwo$ & \cmark & \xmark & \cmark & Security-aware compaction \\
\cmidrule(r){1-1}\cmidrule(r){2-2}\cmidrule(r){3-3}\cmidrule(r){4-4}\cmidrule(r){5-5}\cmidrule(r){6-6}\cmidrule(r){7-7}\cmidrule(r){8-8} 
ARIANNA \cite{collini2025arianna} & eFPGA (bespoke, uniform tiles) & $\rtwo$ & $\rthree$ & \pmark & \cmark & \cmark & 3.3$\times$ lower vs. fixed fabric \\
\cmidrule(r){1-1}\cmidrule(r){2-2}\cmidrule(r){3-3}\cmidrule(r){4-4}\cmidrule(r){5-5}\cmidrule(r){6-6}\cmidrule(r){7-7}\cmidrule(r){8-8} 
SOAR \cite{das2025soar} & eFPGA + Monitoring & $\rfive$ & $\rfour$ & \cmark & \xmark & \cmark & Dynamic Key \\
\cmidrule(r){1-1}\cmidrule(r){2-2}\cmidrule(r){3-3}\cmidrule(r){4-4}\cmidrule(r){5-5}\cmidrule(r){6-6}\cmidrule(r){7-7}\cmidrule(r){8-8} 
TRAP \cite{fowler2025trap} & Transistor-level fabric & $\rone$ & $\rthree$ & \xmark & \xmark & \pmark & SAT-hardening \\
\cmidrule(r){1-1}\cmidrule(r){2-2}\cmidrule(r){3-3}\cmidrule(r){4-4}\cmidrule(r){5-5}\cmidrule(r){6-6}\cmidrule(r){7-7}\cmidrule(r){8-8} 
\multirow{3}{*}{\textbf{NuRedact (proposed)}} & \multirow{3}{*}{eFPGA (non\mbox{-}uniform)} & \multirow{3}{*}{$\rtwo$} & \multirow{3}{*}{$\rthree$} & \multirow{3}{*}{\pmark} & \multirow{3}{*}{\cmark} & \multirow{3}{*}{\cmark} & Low underutilization \\
& & & & & & & Low Black-box Learnability \\
& & & & & & & up to 9x smaller vs. fixed \\
\bottomrule
\end{tabular}
\vspace{3pt}
\raggedright
\scriptsize
Ratings are qualitative and synthesized from reported results in prior work. 
\vspace{-15pt}
\end{table*}

\noindent (i) An automated framework for generating design-specific, \textbf{non-uniform} eFPGA fabrics that reduce area overhead while preserving full functionality of the redacted logic.

\noindent (ii) A Python-based package for layout transformation and VPR architecture customization on a SkyWater 130nm OpenFPGA architecture, for toolchain-compatible fabric generation.

\noindent (iii) A \emph{security-driven overhead-aware selective redaction}, replacing critical logic maximizing overhead reduction with irregular, non-uniform fabrics.

\noindent (iv) Evaluation of overhead using Cadence Genus for synthesis and Synopsys Design Compiler (DC) to produce attack-friendly netlists for evaluation (evaluated by IcySAT unrolling and the KC2 SAT toolchain).

\section{Background and Prior Work}

\subsection{Hardware IP Protection through Redaction}

Prior art in redaction can be divided into three main categories: (i) fine-grain LUT redaction, (ii) coarse-grain eFPGA redaction, and (iii) sub-gate transistor-level redaction:

\noindent (i) \emph{\underline{Fine-grain LUT redaction.}} These techniques redact logic by replacing selected gates/cones with small LUTs \cite{baumgarten2010preventing, kamali2018lut, kolhe2019custom, dasgupta2025hipr, guo2023evolute}. These techniques control overhead by compacting LUTs \cite{kamali2018lut}, using attack-resistive insertion policies \cite{guo2023evolute, dasgupta2025hipr, kamali2018lut}, and bitstream compression \cite{dasgupta2025hipr}. Advantages of these techniques are layout friendliness, medium-effort integration into standard netlists. This is while structure may leak if compaction/placement is careless \cite{dasgupta2025hipr, kolhe2019custom}.

\noindent (ii) \emph{\underline{Fine-grain LUT+Routing redaction.}} Several studies, including HIPR \cite{dasgupta2025hipr}, extend LUT-based redaction by configuring cascaded twisted multiplexers to realize both LUT functionality and routing crossbars \cite{kamali2019full, kamali2020interlock}. This design closely resembles FPGA architectures, in which part of the configuration enables logic operations while the remainder supports routing recovery. Despite its robustness, the approach is vulnerable to link-prediction–based machine-learning attacks \cite{alrahis2021untangle}.

\noindent (iii) \emph{\underline{Coarse-grain eFPGA redaction.}} RTL modules (or IPs) are moved behind an eFPGA macro, where early end-to-end flows automated what to redact \cite{tomajoli2022alice}. More recent frameworks tailor the fabric to each cluster (bespoke K/N/IO/tile grids) after pre-filtering to secure configurations, improving utilization and cutting overhead \cite{collini2025arianna}. These techniques benefit from strong decoupling between revealed structure and hidden behavior and mature tool flows (OpenFPGA/VPR) for full automation. However, they are often over-provisioned for redaction, wasting logic/routing and driving overhead. Additionally, configuration registers and I/O pins can feed attack inferences if, leading to function/bitstream leakage\cite{han2023functeller, karmakar2024evaluating, sathe2023mantis, rezaei2022evaluating}. 

\noindent (iv) \emph{\underline{Sub-gate transistor-level redaction.}} Switch-level fabrics purposefully break gate-level assumptions for overhead mitigation \cite{fowler2025trap}. Similar approaches have been used in the past for only routing-based obfuscation \cite{shamsi2018cross}, where more recent ones shift to full configurable substrate at transistor-level. These approaches eliminate hierarchical LUT cues. However, tooling (automation) is less turnkey than LUT/eFPGA flows. 

Table \ref{tab:background-comparison} shows a detailed comparison of key studies in these three categories. \emph{NuRedact} keeps the integration/automation of eFPGA redaction but discards uniformity. By co-designing non-uniform fabrics with the target cones, it retains the redaction security model while sharply reducing under-utilization and area—precisely the gap revealed by the prior art above.

\subsection{Custom eFPGA Design Frameworks}

An eFPGA redaction flow couples (i) an architecture description (e.g., VTR/VPR XML) that defines grid layout, tiles, logic clusters, routing, etc. (ii) a fabric generator (e.g., OpenFPGA) that binds architectural primitives to RTL and bitstream tooling, and (iii) an ASIC integration path that compiles eFPGA macro with the vendor’s compiler (e.g., Cadence Genus) for timing, floorplaning, DFT, etc. \cite{luu2014vtr, tang2019openfpga}. Various open-source and commercial frameworks have been introduced for the integration of eFPGA into ASICs, i.e., eFPGA-based system-on-chips (SoCs), where VTR/VPR \cite{luu2014vtr} + OpenFPGA \cite{tang2019openfpga} are widely used for prototyping. This tooling stacks provide key knobs for configurations, including logic clustering (LUT size and fracturability, carry chain, and hard blocks), routing topology (e.g., channel width, switch-box types, and crossbar topology), clocking, and configuration protocol (frame-based or memory-based) \cite{tang2019openfpga, kamali2023shell}. 

Crucially, although fabrics are often presented as perfectly tileable and uniform, both VTR/VPR and OpenFPGA natively support heterogeneity and irregularity \cite{luu2014vtr, tang2019openfpga}, including mixing tile types, using irregular pin maps, depopulating connection/switch boxes asymmetrically, and varying channel width. Legality is preserved through pin-equivalence classes, placement regions, and explicit resource constraints, so the standard pack/place/route remains correct even when the fabric deviates from a checkerboard (mesh model) \cite{luu2014vtr, mcmurchie1995pathfinder}. For redaction, these capabilities matter because the objective is not end-user flexibility but a large functional/search space at low silicon cost and with minimal structural regularity. \emph{NuRedact} implements redaction by co-designing the fabric with the target cones using irregular tilings, pin-map asymmetry, and selective logic depopulation, then enforces legality through explicit placer/router constraints and binding metadata, thus retaining security while lowering overhead and integrating cleanly with OpenFPGA/VTR and standard ASIC flows.

\subsection{Attacks and Security Analysis of eFPGA Redaction}

Attacks on redaction aim to (i) recover functionality (partial or full) \cite{han2023functeller, sathe2023mantis} or (ii) shrink the search space until the remaining configuration bits are solvable. \cite{bhandari2023not, rezaei2022evaluating}. For LUT-based schemes, early random LUT insertion fell quickly to SAT de-obfuscation \cite{sweeney2020modeling}. Later heuristics improved hardness but leaked structure, where ML-based attacks target missing k-cuts infer LUT placement/keys \cite{alrahis2021untangle}. For eFPGA-based redaction, the uniformity of the fabric blunts direct structural inference, but black-box functional attacks remain viable, e.g., FuncTeller that queries the eFPGA I/O, identifies on-set minterms, expands them to prime implicants, and synthesizes an approximate netlist \cite{han2023functeller}. In parallel, SAT derivations, e.g., sequential/cyclic SAT (e.g., unrolling), can collapse the effective configuration space once cycles are handled, enabling oracle-guided recovery \cite{zhou2017cycsat, shamsi2019kc2, shamsi2019icysat}. More recently, data-driven approaches such as MANTIS learn approximate bitstreams from observed I/O behavior without an explicit SAT model, underscoring that the attack surface extends beyond classic logic-level formulations \cite{sathe2023mantis}. These trends point to architectural knobs that reduce attack success (targeted by \emph{NuRedact}): (i) limit structural regularity that ML models and oracle-less methods exploit and (ii) raise entropy density by avoiding low-utilization, over-provisioned eFPGA fabrics.

\section{Threat Model}

In line with prior redaction studies, the assets (attack target) would be (i) the functionality of the redacted cones, (ii) the eFPGA configuration (bitstream) that realizes them, and (iii) the mapping between architectural bits and configuration addresses. Attacker goal is to replicate the protected logic by recovering a correct (or approximate) configuration, or by reconstructing the redacted functionality. The attacker has the redacted gate-level netlist of the ASIC and can apply input–output queries to a working device (oracle). Attack surfaces we evaluate is Oracle-guided SAT family, black-box functional recovery, and Oracle-less/structural guidance. We assume bitstreams are not available in plaintext, and configuration interfaces are placed within the system’s security perimeter. Additionally, invasive physical extraction and advanced side-channel attacks on the bitstream loader are important but excluded here to align with prior redaction work.

\section{Proposed Redaction Scheme: \emph{NuRedact}}

\begin{figure}[t]
\centering
\includegraphics[width=\linewidth]{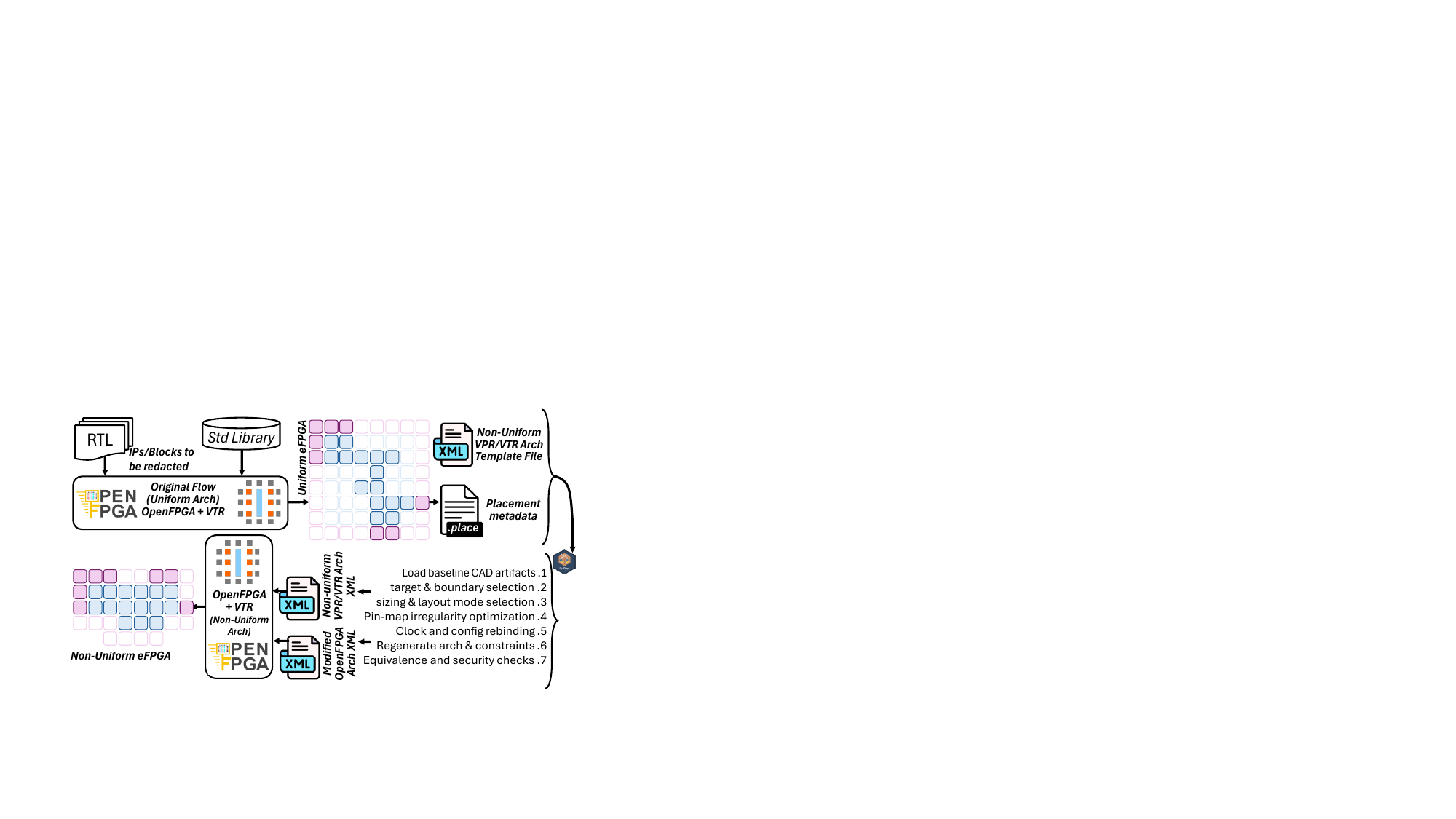}
\caption{\emph{NuRedact} Framework: Automated Generation of Custom Non-Uniform eFPGA Fabrics by Leveraging Placement Data from Uniform Flow and Integrating Python-Based VPR/VTR XML Customization within OpenFPGA.}
\label{fig:NuRedact_flow}
\end{figure}

\emph{NuRedact} is a fully automated flow that synthesizes design-specific, a first-of-its-kind non-uniform eFPGA fabrics for redaction. To guarantee reliable automation (keep everthing running on widely-used frameworks, e.g., OpenFPGA and VPR/VTR), the key idea is to extract structured metadata from the baseline VPR/OpenFPGA placement/routing (from the baseline uniform run), then regenerate a custom layout that preserves function and CAD legality while removing under-utilized logic/routing and injecting architectural irregularity (pin-maps, logic depopulation, regional segments). The result is a compact fabric that matches the redacted cone’s footprint, integrates with the unmodified toolchain, and maintains security at much lower overhead. The overall implementation flow of \emph{NuRedact} is shown in Figure~\ref{fig:NuRedact_flow}, highlighting the integration of our custom package for layout transformation, customized VPR/VTR architecture generation, and OpenFPGA run. 

\subsection{Uniform Fabric Analysis and Logic Utilization Extraction}

\emph{NuRedact} starts the process from a regular (uniform) eFPGA fabric, composed of a regular array of homogeneous logic tiles and I/O pins. The design is packed/placed/routed using the standard OpenFPGA+VPR flow, yielding the canonical artifacts (including \texttt{.place} that records the spatial coordinates of all placed logic blocks and I/O elements). By processing \texttt{.place} log, \emph{NuRedact} extracts a utilization map (occupied tiles and I/Os), a routing map, and timing criticality per net. This analysis enables the elimination of unused logic and I/O resources, initiating customized non-uniform layout that preserves only the actively used tiles and I/Os.










\begin{algorithm}[t]
\footnotesize
\caption{\emph{NuRedact}: Non-Uniform Fabric Generation}
\label{alg:NuRedact}
\begin{algorithmic}[1]

\Input $P$: placement file,\; $A_{temp}$: template arch.
\Output $A_{nu}$: non-uniform arch.

\State $C_{io} \gets$ IO capacity from $A_{temp}$  
\State $(N_{clb}, N_{io}) \gets$ parse($P$)  

\State $B_{io} \gets \left\lceil \tfrac{N_{io}}{C_{io}} \right\rceil$

\State $(W,H) \gets \min_{w,h}(w+2)(h+2)$  
\hskip1em s.t. $wh \geq N_{clb}, \;\; 2(w+h) \geq B_{io}$  

\State Place $N_{clb}$ CLBs $\to \mathcal{C}$  
\State Boundary IOs $\to \mathcal{I}$  

\While{$|\mathcal{I}| < B_{io}$}  
    \State $\mathcal{I} \gets \mathcal{I} \cup \mathcal{P}$  
\EndWhile  

\State $A_{nu} \gets \mathcal{C} \cup \mathcal{I}$  
\State Export $A_{nu} \to$ VPR XML  

\end{algorithmic}
\end{algorithm}

\subsection{Automated Generation of Non-Uniform Layouts}

Given the placement data, a Python-based layout transformation framework is implemented in \emph{NuRedact} that constructs a customized non-uniform architecture (w.r.t. structured metadata from \texttt{.place} log). Based on two inputs, (i) structured metadata from \texttt{.place} log, and (ii) parameterized VPR architecture template XML, this Python-based framework generates a new architecture XML file with an updated layout and routing configuration that reflects a reduced and design-specific layout while maintaining compatibility with VPR's input requirements, as illustrated in Algorithm \ref{alg:NuRedact}.

\begin{figure}[t]
\centering
\includegraphics[width=\linewidth]{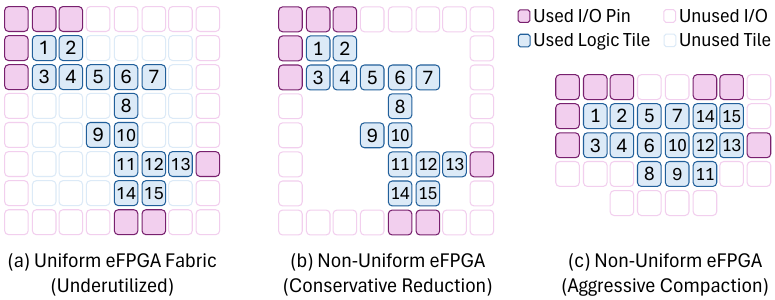}
\caption{Uniform and Non-Uniform eFPGAs in \emph{NuRedact} using (b) Conservative Redaction and (c) Aggressive Layout Compaction.}
\label{fig:uniform_nonuniform_arch}
\vspace{-5pt}
\end{figure}

Unlike conventional VPR architectures that utilize \texttt{auto\_layout} to define tile arrangement, the customized non-uniform architectures rely on a \textit{fixed\_layout} specification. This allows for the explicit definition of tile locations, enabling precise control over the spatial organization of logic tiles and I/O pins. The overall array size in the fixed layout is not static but instead dynamically determined by the spatial extent of the utilized tiles in the original placement. As a result, the final fabric dimensions directly reflect the design’s actual resource demands. Our framework supports two layout strategies: 

\noindent (i) \emph{\underline{Conservative Reduction}}: It preserves the original spatial coordinates of all utilized logic tiles and removes all unused tiles. I/O tiles are retained only up to the required number, with at least one I/O per side to maintain routing accessibility. This approach ensures minimal disruption to routing and timing while achieving resource reduction. An example of this strategy is shown in Figure~\ref{fig:uniform_nonuniform_arch}(b). 

\noindent (ii) \emph{\underline{Layout Compaction}}: It applies aggressive compaction by relocating all utilized logic tiles into a contiguous, tightly packed region. This relocation minimizes the overall fabric dimensions and improves spatial locality. I/O tiles are selected based on adjacency to used logic tiles, and the number of I/O tiles is reduced by adjusting per-tile I/O capacity where applicable. An example of this strategy is shown in Figure~\ref{fig:uniform_nonuniform_arch}(c). To support flexible routing in the resulting compact layouts, \emph{NuRedact} uses an additional routing segment into the customized VPR XML. In particular, a short-length segment (L1) is introduced alongside the default long-length segment (L4). While L4 spans four logic blocks and eases longer communication across the fabric, the newly added L1 segment is limited to a single block length. This finer routing improves local connectivity and is particularly beneficial in densely packed, non-uniform layouts where short interconnects dominate. The combination of L1 and L4 routing resources enhances routability and mitigates congestion that may arise due to irregular tile spacing.

\subsection{Fabric Regeneration and Flow Compatibility}

The custom architecture file produced by the transformation script is used to regenerate the fabric through a second invocation of the OpenFPGA flow. This stage produces the non-uniform eFPGA fabric, including routing infrastructure and configuration bitstream generation support. The regenerated architecture remains compatible with the original RTL design and synthesis results. The process preserves all logical mappings and ensures consistent functionality, while delivering a reduced-area implementation. Integration into OpenFPGA's task-based infrastructure allows the entire process, starting from a uniform run and ending with a synthesized non-uniform fabric, to be executed automatically without manual intervention. \emph{NuRedact} enables the construction of logic-agnostic, general-purpose non-uniform eFPGA fabrics that are tightly matched to actual design requirements. By eliminating unused tiles and compacting placement, the resulting architectures demonstrate improved spatial efficiency and reduced overhead without sacrificing correctness or tool compatibility.

\section{Secure IP Redaction Using \textit{NuRedact} Framework}

From the redaction standpoint, \textit{NuRedact} introduces spatially constrained intra-IP redaction that eases selective redaction of critical sub-components with low external observability, e.g., arithmetic blocks of encryption cores, control logic, or decision-making units. As \textit{NuRedact} maximizes the utilization ratio per logic tile (using logic depopulation), the ratio of logic to I/O would be increased drastically (denser logic in a smaller footprint and fewer I/O pins). Accordingly, this non-uniform redaction, coupled with pin-map asymmetry, minimizes black-box learnability by algorithmic attacks, e.g., SAT-derived attacks. Our experiments (see Section \ref{sec:deobfuscation}) demonstrate that this dense logic in a smaller footprint increases SAT variables per bitstream size by one order of magnitude, showcasing the security enhancement at lower PPA overhead.

Additionally, in \textit{NuRedact}, ASIC to/from eFPGA boundary is chosen by a simple multi-objective policy: minimize interface width (I/O cut) and exposed controllability/observability, subject to timing slack and floorplan constraints. In practice, we rank candidate cuts by (i) cut-signal count and toggle/observe metrics, (ii) uniqueness/rarity of the logic, and (iii) a detour budget for top-critical nets. Architectural irregularity is introduced by pin-map asymmetry (permuting equivalent pins to break port regularity) and logic/routing depopulation to further strengthen security by eliminating predictable architectural patterns. Compared to square, uniform overlays, this (i) removes under-utilized silicon, (ii) reduces visual and graph-structural cues that aid oracle-less inference, and (iii) raises configuration entropy per logic, all while running through standard OpenFPGA/VTR and sign-off flows. 

\section{Experimental Results and analysis}

\begin{table}[t]
\footnotesize
\centering
\setlength{\tabcolsep}{5pt}
\caption{Specifications of the selected Benchmark Circuits.}
\label{tab:benchmark}
\begin{tabular}{@{} l *{21}c @{}}
\toprule
\textbf{Benchmark} & \textbf{\# Modules} & \textbf{\# Inputs} & \textbf{\# Outputs} & \textbf{Redacted Modules}\\

\cmidrule(r){1-1}\cmidrule(r){2-2}\cmidrule(r){3-3}\cmidrule(r){4-4}\cmidrule(r){5-5}

GPS & 12 & 6-128 & 1-256 & \texttt{Cacode}  \\
\cmidrule(r){1-1}\cmidrule(r){2-2}\cmidrule(r){3-3}\cmidrule(r){4-4}\cmidrule(r){5-5}

\multirow{3}{*}{RISC-V} & \multirow{3}{*}{18} & \multirow{3}{*}{3-130} & \multirow{3}{*}{1-79} & \texttt{CTRL}  \\
 & & & & \texttt{Arbiter} \\
  & & & & \texttt{Ld/St} \\
\cmidrule(r){1-1}\cmidrule(r){2-2}\cmidrule(r){3-3}\cmidrule(r){4-4}\cmidrule(r){5-5}

DES3 & 11 & 6-240 & 4-64 & \texttt{SBOX\_8} \\
\cmidrule(r){1-1}\cmidrule(r){2-2}\cmidrule(r){3-3}\cmidrule(r){4-4}\cmidrule(r){5-5}

AES & 9 & 10-128 & 8-128 & \texttt{AES\_ShR} \\
\cmidrule(r){1-1}\cmidrule(r){2-2}\cmidrule(r){3-3}\cmidrule(r){4-4}\cmidrule(r){5-5}

\multirow{2}{*}{ADDER} & \multirow{2}{*}{2} & \multirow{2}{*}{3-4} & \multirow{2}{*}{2-3} & \texttt{FA\_ARRAY} \\
&  &  &  & \texttt{HA\_ARRAY} \\
\cmidrule(r){1-1}\cmidrule(r){2-2}\cmidrule(r){3-3}\cmidrule(r){4-4}\cmidrule(r){5-5}

GCD & 8 & 8-45 & 1-18 & \texttt{Comparator} \\

\bottomrule

\end{tabular}
\end{table}

To evaluate the effectiveness of the proposed \emph{NuRedact} framework, we conducted experiments on a set of benchmarks, including a RISC-V–based PULPino SoC and additional IP cores listed in Table \ref{tab:benchmark}. For each benchmark, as shown, specific modules were selected as targets for eFPGA-based redaction using the OpenFPGA toolchain\footnote{Amongst the Benchmarks, smaller ranges are selected to visualize the compactness effectiveness of \emph{NuRedact} over the fabrics.}. \emph{NuRedact} is implemented in Python and leverages the PyVerilog for (lexical) parsing and generation of non-uniform eFPGAs, all integrated into the OpenFPGA flow. Logic synthesis was performed using Cadence Genus 21.18 with the Skywater 130nm technology library to evaluate area overheads, on a server with an Intel Xeon E5-2640 v4 2.4 GHz processor (40 cores, 48 GB RAM). To assess the security of the redacted fabrics, netlists compatible with IcySAT and KC2 SAT attack tools were generated using Synopsys DC on Nangate 15nm technology node. Attacks were launched on the resulting non-uniform eFPGA fabrics using an Ubuntu environment on an Intel® Core™ i7-14700 system (28 cores, 32 GB RAM).

\begin{table}[b]
\scriptsize
\centering
\setlength{\tabcolsep}{2pt}
\caption{Area Efficiency of eFPGA Fabrics Generated via the \emph{NuRedact} Framework Compared to Traditional Uniform Architectures.}
\label{tab:uniform_nonuniform_area}
\begin{tabular}{c c c c c cc c}
\toprule
\multirow{2}{*}{\textbf{IP}} & \multirow{2}{*}{\textbf{Module}} & \multirow{2}{*}{\textbf{Fabric}} & \multirow{2}{*}{\textbf{\# Total }} & \multirow{2}{*}{\textbf{\# Used }} & \multicolumn{2}{c}{\textbf{Area(($\mu$m$^2$)}} & \multirow{2}{*}{\textbf{$\Delta \text{Area}$}}\\
\cmidrule(r){6-7}
& & \textbf{Size} & \textbf{Tiles} & \textbf{Tiles} & \textbf{Uniform} & \textbf{Non-Uni} &  (\%)\\
\cmidrule(r){1-1}\cmidrule(r){2-2}\cmidrule(r){3-3}\cmidrule(r){4-4}\cmidrule(r){5-5}\cmidrule(r){6-6}\cmidrule(r){7-7}\cmidrule(r){8-8}

GPS & \texttt{Cacode} & 3x3 & 9 & 7 & 513,361 & 357,251 & -30.41 \\
\cmidrule(r){1-1}\cmidrule(r){2-2}\cmidrule(r){3-3}\cmidrule(r){4-4}\cmidrule(r){5-5}\cmidrule(r){6-6}\cmidrule(r){7-7}\cmidrule(r){8-8}

AES & \texttt{AES\_Shr} & 2x2 & 4 & 3 & 212,716 & 158,113 & -25.67 \\ 
\cmidrule(r){1-1}\cmidrule(r){2-2}\cmidrule(r){3-3}\cmidrule(r){4-4}\cmidrule(r){5-5}\cmidrule(r){6-6}\cmidrule(r){7-7}\cmidrule(r){8-8}

Seq\_Comb & \texttt{Seq\_Comb} & 3x3 & 9 & 5 & 560,845 & 257,653.5 & -54.06  \\
\cmidrule(r){1-1}\cmidrule(r){2-2}\cmidrule(r){3-3}\cmidrule(r){4-4}\cmidrule(r){5-5}\cmidrule(r){6-6}\cmidrule(r){7-7}\cmidrule(r){8-8}

\multirow{3}{*}{RISC-V} & \texttt{CTRL} & 3x3 & 9 & 5 & 529,113 & 306,957 & -41.99 \\
\cmidrule(r){2-2}\cmidrule(r){3-3}\cmidrule(r){4-4}\cmidrule(r){5-5}\cmidrule(r){6-6}\cmidrule(r){7-7}\cmidrule(r){8-8}

& \texttt{Arbiter} & 6x6 & 36 & 5 & 2026,867 & 353,273.5 & -82.57 \\
\cmidrule(r){2-2}\cmidrule(r){3-3}\cmidrule(r){4-4}\cmidrule(r){5-5}\cmidrule(r){6-6}\cmidrule(r){7-7}\cmidrule(r){8-8}

& \texttt{Ld/St} & 1x1 & 1 & 1 & 64,195 & 57,833 & -9.91 \\
\cmidrule(r){1-1}\cmidrule(r){2-2}\cmidrule(r){3-3}\cmidrule(r){4-4}\cmidrule(r){5-5}\cmidrule(r){6-6}\cmidrule(r){7-7}\cmidrule(r){8-8}

Logic7 & \texttt{Logic7} & 3x3 & 9 & 7 & 513,361 & 357,251 & -30.41 \\
\cmidrule(r){1-1}\cmidrule(r){2-2}\cmidrule(r){3-3}\cmidrule(r){4-4}\cmidrule(r){5-5}\cmidrule(r){6-6}\cmidrule(r){7-7}\cmidrule(r){8-8}

ADDER & \texttt{FA\_Array} & 1x1 & 1 & 1 & 64,195 & 57,833 & -9.91 \\
\cmidrule(r){1-1}\cmidrule(r){2-2}\cmidrule(r){3-3}\cmidrule(r){4-4}\cmidrule(r){5-5}\cmidrule(r){6-6}\cmidrule(r){7-7}\cmidrule(r){8-8}

DES3 & \texttt{SBOX\_8} & 1x1 & 1 & 1 & 64,195 & 57,833 & -9.91 \\

\bottomrule
\multicolumn{8}{l}{ Negative values of $\Delta \text{Area}$ (\%) indicate area reduction relative to the uniform fabric.}

\end{tabular}
\end{table}

\subsection{Hardware Overhead Analysis}

\begin{figure}[t]
\centering
\includegraphics[width=\linewidth]{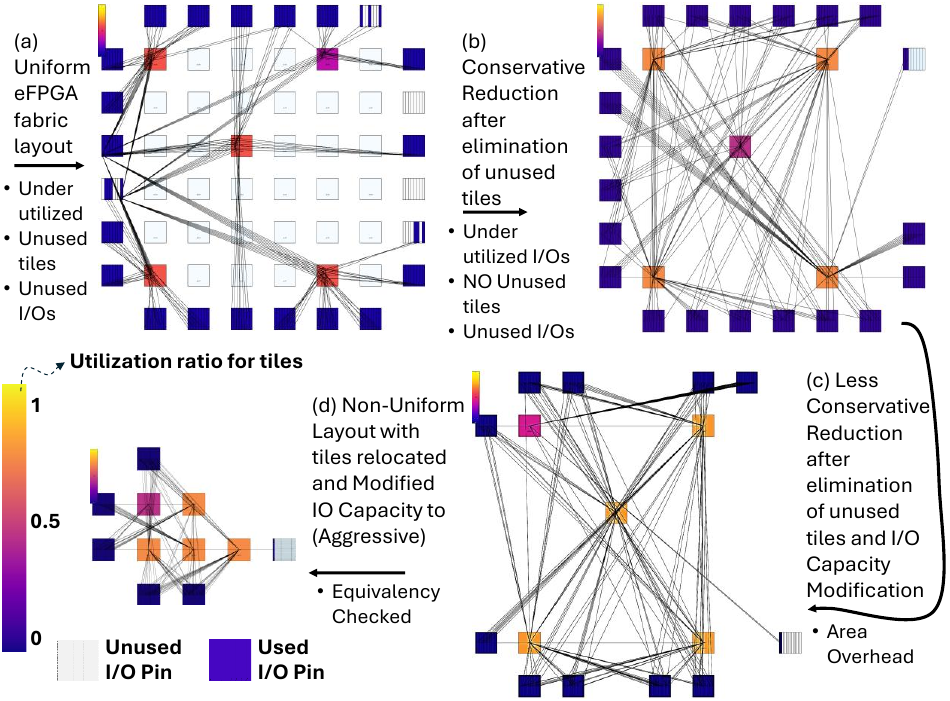}
\caption{Comparison of uniform and non-uniform eFPGA fabric layouts for \texttt{Arbiter} module of the RISC-V benchmark, generated using the \emph{NuRedact} framework (using both Conservative and Aggressive).}
\label{fig:layout_arbiter}
\end{figure}

\begin{figure}[t]
\centering
\includegraphics[width=\linewidth]{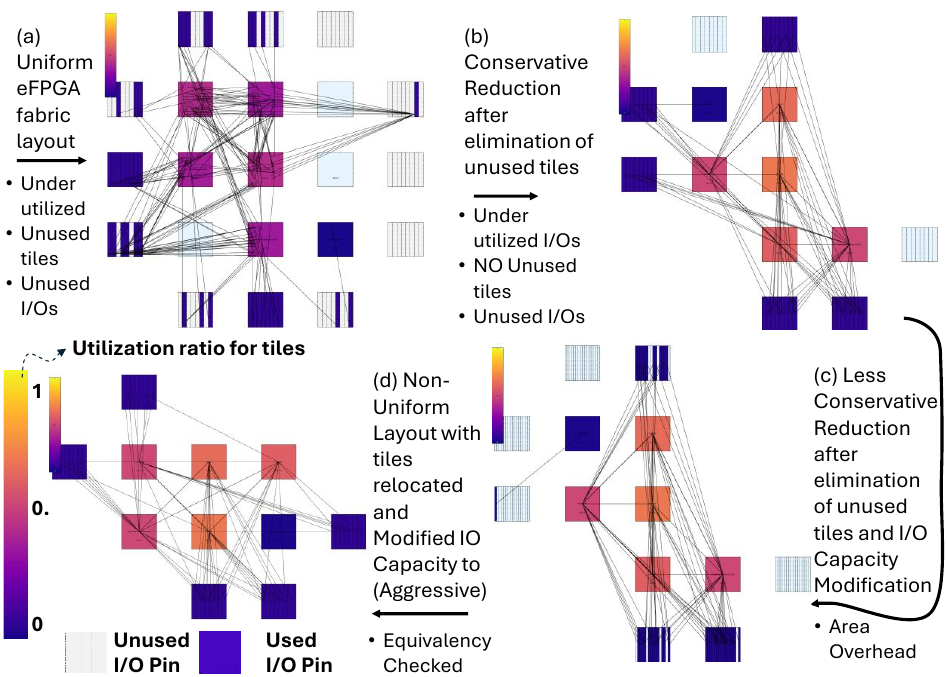}
\caption{Comparison of uniform and non-uniform eFPGA fabric layouts for \texttt{CTRL} module of the RISC-V benchmark, generated using the \emph{NuRedact} framework (using both Conservative and Aggressive)}
\label{fig:layout_controller}
\end{figure}

The primary goal of \emph{NuRedact} framework is to minimize eFPGA fabric area while preserving functionality and robustness against attacks. For each benchmark, both uniform (baseline) and non-uniform fabrics are generated by the \emph{NuRedact} flow. Fabric layouts were obtained with VPR, leveraging a custom VPR architecture file together with the BLIF representation of the target modules. Figures \ref{fig:layout_arbiter} and \ref{fig:layout_controller} illustrate the transformation pipeline on two RISC-V modules, from uniform to compacted non-uniform, all extracted from OpenFPGA framework. For  \texttt{Arbiter} (Figure \ref{fig:layout_arbiter}(a)) is a 6x6 fabric, while after removing unused CLBs, a reduced non-uniform layout is obtained (Figure \ref{fig:layout_arbiter}(b)), further refined by the conservative reduction (Figure \ref{fig:layout_arbiter}(c)), where unused CLBs are pruned and the I/O capacity is adjusted. This is followed by compact layout generation (Figure \ref{fig:layout_arbiter}(d)), where the I/O capacity is increased and the used tiles are relocated into a contiguous region to achieve a more compact fabric. This sequence shrinks the footprint of \texttt{Arbiter} from 6×6 tiles to a five-tile design ($\sim$86\% or 7.14x area reduction), while keeping the standard tool flow unchanged.

The \texttt{CTRL} shows the same trend (Figure \ref{fig:layout_controller}), albeit with a smaller compaction ratio due to a more fit logic footprint. Heatmaps by OpenFPGA confirm that \emph{NuRedact} increases per-tile utilization: for the \texttt{Arbiter}, the average tile utilization rises from 0.68 (uniform) to 0.94 (aggressive compaction). Higher utilization yields fewer idle configuration bits and a denser configuration entropy per unit area, which reduces structural cues and limits black-box learnability, while preserving CAD legality and compatibility with OpenFPGA/VTR.

\begin{table}[b]
\scriptsize
\centering
\setlength{\tabcolsep}{2pt}
\caption{{Area Optimization in \textit{\uppercase{N}u\uppercase{A}rch}: Conservative vs. Aggressive.}}
\label{tab:two_nonuniform_area_comp}
\begin{tabular}{c c cc cc c}
\toprule
\multirow{3}{*}{\textbf{Module}} & \multirow{3}{*}{\textbf{\#IO/Tile}} &
\multicolumn{2}{c}{\textbf{Conservative}} &
\multicolumn{2}{c}{\textbf{Aggressive}} &
\multirow{3}{*}{$\boldsymbol{\Delta}$\textbf{Area} (\%)} \\
\cmidrule(r){3-4} \cmidrule(r){5-6}
& & \#I/O & Area  & \#I/O Required & Area  & \\
& & Tile & ($\mu$m$^2$) & (Allocated)  & ($\mu$m$^2$) & \\
\midrule  

\multirow{3}{*}{\texttt{Seq\_Comb}} & 4            & 6  & 264,708 & 6    & 255,618   & -3.43  \\
                          & 8 (Default)   & 6  & 270,975 & 3(4) & 257,653.5 & -4.92 \\
                          & 16            & 6  & 280,724 & 2(4) & 264,199   & -5.89 \\
\midrule

\multirow{3}{*}{\texttt{CTRL}} & 4           & 10 & 315,753.5 & 10   & 307,256.5 & -2.69  \\
                          & 8 (Default)   & 7  & 319,507   & 5    & 306,957   & -3.93 \\
                          & 16            & 7  & 331,092.5  & 2(4) & 311,303   & -5.98 \\
\midrule

\multirow{4}{*}{\texttt{Arbiter}} & 8 (Default)  & 21 & 384,266   & 21 & 353,273.5 & -8.06  \\
                          & 16            & 11 & 381,784.5 & 11 & 321,403   & -15.82 \\
                          & 32            & 6 & 410,700    & 6  & 325,971   & -20.63 \\
                          & 48            & 6 & 527,432.5  & 4  & 327,645.5 & -37.88 \\
\midrule
\multirow{3}{*}{\texttt{Logic7}} & 4              & 10  & 363,347.5  & 6    & 355,217.5  & -2.24  \\
                          & 8 (Default)  & 10  & 373,705.5  & 4    & 357,251    & -4.40 \\
                          & 16           & 10  & 389,769.5  & 2(4) & 363,775.55 & -6.67 \\

\bottomrule
\multicolumn{6}{l}{Negative values of $\Delta \text{Area}$ (\%) indicate area reduction.}

\end{tabular}
\end{table}

We synthesized both the uniform and \emph{NuRedact} non-uniform fabrics with Cadence Genus under identical libraries and timing constraints to compare silicon area. Table \ref{tab:uniform_nonuniform_area} summarizes results across all redacted modules. As an instance, for \texttt{Seq\_Comb} (3×3 fabric with 5 used tiles), \emph{NuRedact} cuts area by 54.06\% (2.18x) relative to the uniform baseline, and for the RISC-V \texttt{Arbiter} (baseline 6×6), the reduction reaches 82.57\%\footnote{~Per tile area reduction estimated from OpenFPGA layout counts is $\approx 86\%$, whereas post-synthesis area from Cadence Genus reports $82.57\%$; the close agreement validates the per-tile comparison as a quick proxy for silicon area.} (5.73x), highlighting the efficiency of tailoring the fabric footprint to actual utilization.

Table \ref{tab:two_nonuniform_area_comp} compares the two non-uniform modes, conservative reduction vs. layout compaction, while sweeping the per-tile I/O capacity (e.g., 4, 8, 16, 32; and 48 for \texttt{Arbiter}). To preserve routability, we kept at least one I/O tile on each fabric edge even when demand fell below four. The general trend is consistent: as per-tile I/O capacity increases, \emph{NuRedact}’s layout compaction yields larger area savings over conservative reduction. For example, for larger modules like \texttt{Arbiter}, with higher I/Os, aggressive mode achieves 37.88\% area reduction for compaction vs. conservative (Total of 89.17\% (9.23x reduction) vs. uniform. Across all benchmarks, layout compaction produces the most compact fabrics (smallest area), while conservative reduction provides a lower-risk fallback when timing/routability are tight.

\subsection{De-obfuscation Analysis} \label{sec:deobfuscation}

For the security analysis, we redacted each target module onto two fabrics: (i) the standard SkyWater-130 OpenFPGA architecture, and (ii) the non-uniform VPR architecture generated by \emph{NuRedact}. Both fabrics use scan-chain configuration FFs and fracturable 4-input LUTs. Similar to prior state-of-the-art eFPGA-based (and other reconfigurable-based) studies \cite{collini2025arianna, karmakar2024evaluating, dasgupta2025hipr, fowler2025trap}, our security analysis of \emph{NuRedact} will focus exclusively on derived SAT attacks. This is because eFPGA-based redaction is inherently resilient against other categories of de-obfuscation attacks, such as structural- or ML-based approaches \cite{karmakar2024evaluating, rezaei2022evaluating}. To enable SAT-style analysis (for both cyclic and sequential), we convert the redacted fabrics to \texttt{.bench} by exposing all configuration FFs as primary key inputs. The configuration FFs are stitched as a scan chain driven by \texttt{prog\_clk} (we recover the correct bit ordering via a depth-first walk from \texttt{scan\_in\_head} to \texttt{scan\_in\_tail}), then validate the mapping by re-simulation against the OpenFPGA bitstream. For sequential/cyclic behavior, we use IcySAT to unroll hard combinational loops and instantiate an oracle by fixing the key inputs to the ground-truth configuration from the OpenFPGA bitstream. We then launch KC2 on the key-exposed netlist to assess functional recovery. (Time/memory budgets are fixed across all runs; uniform vs. non-uniform are evaluated under identical conditions.)

Table \ref{tab:SAT_Attack_NuRedact} summarizes SAT outcomes on \emph{NuRedact}ed fabrics. For a small design (2-bit adder, bitstream size 915), IcySAT unrolled 98 loops and KC2 recovered the key in 2+ hours. Compared to its uniform counterpart, the non-uniform fabrics introduced $\ge$10× SAT variables on \emph{NuRedact}ed modules, substantially increasing solve time, which is due to reducing structural cues and black-box learnability. For larger designs and bitstreams (e.g., the RISC-V CTRL and GPS Cacode), IcySAT completed unrolling after a long runtime, but KC2 did not recover a key within the available RAM and time budgets.

\begin{table}[t]
\scriptsize
\centering
\setlength{\tabcolsep}{1.2pt}
\caption{Results of SAT-Attack on Non-uniform eFPGA Fabrics Generated via the \emph{NuRedact} Framework.}
\label{tab:SAT_Attack_NuRedact}
\begin{tabular}{@{} l *{21}c @{}}
\toprule
{\textbf{Design}} & {\textbf{Bitstream}} & {\textbf{Unroll Factor}} & {\textbf{\# Clauses}} & {\textbf{\# Variables}} & {\textbf{Time (s) }} & {\textbf{Key (S/F) }}\\

\midrule

\texttt{HA\_ARRAY} & 915 & 98 & 6165189 & 2318300 & 8102.2 & S  \\
\cmidrule(r){1-1}\cmidrule(r){2-2}\cmidrule(r){3-3}\cmidrule(r){4-4}\cmidrule(r){5-5}\cmidrule(r){6-6}\cmidrule(r){7-7}

\texttt{CTRL} & 8215 & 434 & - & - & TO & F \\
\cmidrule(r){1-1}\cmidrule(r){2-2}\cmidrule(r){3-3}\cmidrule(r){4-4}\cmidrule(r){5-5}\cmidrule(r){6-6}\cmidrule(r){7-7}

\texttt{FA\_ARRAY} & 937 & 254 & 19595603 & 7327291 & 40818.5 & S \\
\cmidrule(r){1-1}\cmidrule(r){2-2}\cmidrule(r){3-3}\cmidrule(r){4-4}\cmidrule(r){5-5}\cmidrule(r){6-6}\cmidrule(r){7-7}

\texttt{Cacode} & 9582 & 534 & - & - & TO & F \\
\cmidrule(r){1-1}\cmidrule(r){2-2}\cmidrule(r){3-3}\cmidrule(r){4-4}\cmidrule(r){5-5}\cmidrule(r){6-6}\cmidrule(r){7-7}

\texttt{Ld/St} & 937 & 254 & - & - & TO & F \\
\cmidrule(r){1-1}\cmidrule(r){2-2}\cmidrule(r){3-3}\cmidrule(r){4-4}\cmidrule(r){5-5}\cmidrule(r){6-6}\cmidrule(r){7-7}

\texttt{Comparator} & 861 & 303 & - & - & TO & F \\
\cmidrule(r){1-1}\cmidrule(r){2-2}\cmidrule(r){3-3}\cmidrule(r){4-4}\cmidrule(r){5-5}\cmidrule(r){6-6}\cmidrule(r){7-7}

\texttt{AES\_ShR} & 2252 & 813 & - & - & TO & F \\
\cmidrule(r){1-1}\cmidrule(r){2-2}\cmidrule(r){3-3}\cmidrule(r){4-4}\cmidrule(r){5-5}\cmidrule(r){6-6}\cmidrule(r){7-7}

\texttt{Arbiter} & 8278 & - & - & - & TO & F \\
\cmidrule(r){1-1}\cmidrule(r){2-2}\cmidrule(r){3-3}\cmidrule(r){4-4}\cmidrule(r){5-5}\cmidrule(r){6-6}\cmidrule(r){7-7}

\texttt{SBOX\_8} & 2001 & - & - & - & TO & F\\
\cmidrule(r){1-1}\cmidrule(r){2-2}\cmidrule(r){3-3}\cmidrule(r){4-4}\cmidrule(r){5-5}\cmidrule(r){6-6}\cmidrule(r){7-7}

\texttt{Logic7} & 4787 & - & - & - & TO & F  \\

\bottomrule

\end{tabular}
\end{table}

\section{Conclusion}

This paper introduced \emph{NuRedact}, an automated eFPGA-based redaction flow that replaces security-critical logic with design-specific, non-uniform eFPGA fabrics. \emph{NuRedact} collapses eFPGA under-utilization by injecting pin-map asymmetry, selective logic depopulation, local routing, and compact fixed layout, while remaining CAD-legal compatible with OpenFPGA/VTR and standard sign-off. On RISC-V and other benchmarks, \emph{NuRedact} delivers substantial efficiency; up to 9x area reduction, while per-tile utilization improves by 38\% on average, which helps increase configuration-entropy density per unit area for reducing structural cues and black-box learnability. In (cyclic and sequential) attack evaluations, very small fabrics remain solvable, but non-uniform fabrics with $\ge$5 logic tiles provide practical resistance against de-obfuscation attacks (variables to resolve is increased by 10x).

\bibliographystyle{IEEEtran}
\bibliography{refs}

@article{rostami2014primer,
  title={A primer on hardware security: Models, methods, and metrics},
  author={Rostami, Masoud and Koushanfar, Farinaz and Karri, Ramesh},
  journal={Proceedings of the IEEE},
  volume={102},
  number={8},
  pages={1283--1295},
  year={2014},
  publisher={IEEE}
}

@inproceedings{subramanyan2015evaluating,
  title={Evaluating the security of logic encryption algorithms},
  author={Subramanyan, Pramod and Ray, Sayak and Malik, Sharad},
  booktitle={2015 IEEE International Symposium on Hardware Oriented Security and Trust (HOST)},
  pages={137--143},
  year={2015},
  organization={IEEE}
}

@article{azar2019smt,
  title={SMT attack: Next generation attack on obfuscated circuits with capabilities and performance beyond the SAT attacks},
  author={Azar, Kimia Zamiri and Kamali, Hadi Mardani and Homayoun, Houman and Sasan, Avesta},
  journal={IACR Transactions on Cryptographic Hardware and Embedded Systems},
  pages={97--122},
  year={2019}
}

@article{barenghi2012fault,
  title={Fault injection attacks on cryptographic devices: Theory, practice, and countermeasures},
  author={Barenghi, Alessandro and Breveglieri, Luca and Koren, Israel and Naccache, David},
  journal={Proceedings of the IEEE},
  volume={100},
  number={11},
  pages={3056--3076},
  year={2012},
  publisher={IEEE}
}

@inproceedings{kocher1999differential,
  title={Differential power analysis},
  author={Kocher, Paul and Jaffe, Joshua and Jun, Benjamin},
  booktitle={Annual international cryptology conference},
  pages={388--397},
  year={1999},
  organization={Springer}
}

@article{kamali2022advances,
  title={Advances in logic locking: Past, present, and prospects},
  author={Kamali, Hadi Mardani and Azar, Kimia Zamiri and Farahmandi, Farimah and Tehranipoor, Mark},
  journal={Cryptology ePrint Archive},
  year={2022}
}

@inproceedings{collini2022reconfigurable,
  title={Reconfigurable logic for hardware IP protection: Opportunities and challenges},
  author={Collini, Luca and Tan, Benjamin and Pilato, Christian and Karri, Ramesh},
  booktitle={Proceedings of the 41st IEEE/ACM International Conference on Computer-Aided Design},
  pages={1--7},
  year={2022}
}

@article{abideen2024overview,
  title={An overview of FPGA-inspired obfuscation techniques},
  author={Abideen, Zain Ul and Gokulanathan, Sumathi and J. Aljafar, Muayad and Pagliarini, Samuel},
  journal={ACM Computing Surveys},
  volume={56},
  number={12},
  pages={1--35},
  year={2024},
  publisher={ACM New York, NY}
}

@inproceedings{kamali2018lut,
  title={Lut-lock: A novel lut-based logic obfuscation for fpga-bitstream and asic-hardware protection},
  author={Kamali, Hadi Mardani and Azar, Kimia Zamiri and Gaj, Kris and Homayoun, Houman and Sasan, Avesta},
  booktitle={2018 IEEE Computer Society Annual Symposium on VLSI (ISVLSI)},
  pages={405--410},
  year={2018},
  organization={IEEE}
}

@inproceedings{bhandari2021exploring,
  title={Exploring eFPGA-based redaction for IP protection},
  author={Bhandari, Jitendra and Moosa, Abdul Khader Thalakkattu and Tan, Benjamin and Pilato, Christian and Gore, Ganesh and Tang, Xifan and Temple, Scott and Gaillardon, Pierre-Emmanuel and Karri, Ramesh},
  booktitle={2021 IEEE/ACM International Conference On Computer Aided Design (ICCAD)},
  pages={1--9},
  year={2021},
  organization={IEEE}
}

@inproceedings{tomajoli2022alice,
  title={ALICE: An automatic design flow for eFPGA redaction},
  author={Tomajoli, Chiara Muscari and Collini, Luca and Bhandari, Jitendra and Moosa, Abdul Khader Thalakkattu and Tan, Benjamin and Tang, Xifan and Gaillardon, Pierre-Emmanuel and Karri, Ramesh and Pilato, Christian},
  booktitle={Proceedings of the 59th ACM/IEEE Design Automation Conference},
  pages={781--786},
  year={2022}
}

@article{dasgupta2025hipr,
  title={HIPR: Hardware IP Protection through Low-Overhead Fine-Grain Redaction},
  author={Dasgupta, Aritra and Paria, Sudipta and Bhunia, Swarup},
  journal={IACR Transactions on Cryptographic Hardware and Embedded Systems},
  volume={2025},
  number={3},
  pages={781--805},
  year={2025}
}

@article{bhandari2023not,
  title={Not all fabrics are created equal: Exploring eFPGA parameters for IP redaction},
  author={Bhandari, Jitendra and Moosa, Abdul Khader Thalakkattu and Tan, Benjamin and Pilato, Christian and Gore, Ganesh and Tang, Xifan and Temple, Scott and Gaillardon, Pierre-Emmanuel and Karri, Ramesh},
  journal={IEEE Transactions on Very Large Scale Integration (VLSI) Systems},
  volume={31},
  number={10},
  pages={1459--1471},
  year={2023},
  publisher={IEEE}
}

@article{collini2025arianna,
  title={ARIANNA: An Automatic Design Flow for Fabric Customization and eFPGA Redaction},
  author={Collini, Luca and Bhandari, Jitendra and Muscari Tomajoli, Chiara and Moosa, Abdul and Tan, Benjamin and Tang, Xifan and Gaillardon, Pierre-Emanuel and Karri, Ramesh and Pilato, Christian},
  journal={ACM Transactions on Design Automation of Electronic Systems},
  year={2025},
  publisher={ACM New York, NY}
}

@article{fowler2025trap,
  title={A TRAP for SAT: On the Imperviousness of a Transistor-Level Programmable Fabric to Satisfiability-Based Attacks},
  author={Fowler, Aric and Mohammed, Shayan and Shihab, Mustafa and Broadfoot, Thomas and Beerel, Peter and Sechen, Carl and Makris, Yiorgos},
  journal={IACR Transactions on Cryptographic Hardware and Embedded Systems},
  volume={2025},
  number={2},
  pages={579--603},
  year={2025}
}

@article{el2019sat,
  title={The SAT attack on IC camouflaging: Impact and potential countermeasures},
  author={El Massad, Mohamed and Garg, Siddharth and Tripunitara, Mahesh V},
  journal={IEEE Transactions on Computer-Aided Design of Integrated Circuits and Systems},
  volume={39},
  number={8},
  pages={1577--1590},
  year={2019},
  publisher={IEEE}
}

@inproceedings{shamsi2017appsat,
  title={AppSAT: Approximately deobfuscating integrated circuits},
  author={Shamsi, Kaveh and Li, Meng and Meade, Travis and Zhao, Zheng and Pan, David Z and Jin, Yier},
  booktitle={2017 IEEE International Symposium on Hardware Oriented Security and Trust (HOST)},
  pages={95--100},
  year={2017},
  organization={IEEE}
}

@inproceedings{shen2017double,
  title={Double DIP: Re-evaluating security of logic encryption algorithms},
  author={Shen, Yuanqi and Zhou, Hai},
  booktitle={Proceedings of the Great Lakes Symposium on VLSI 2017},
  pages={179--184},
  year={2017}
}

@inproceedings{zhou2017cycsat,
  title={CycSAT: SAT-based attack on cyclic logic encryptions},
  author={Zhou, Hai and Jiang, Ruifeng and Kong, Shuyu},
  booktitle={2017 IEEE/ACM International Conference on Computer-Aided Design (ICCAD)},
  pages={49--56},
  year={2017},
  organization={IEEE}
}

@inproceedings{shamsi2019icysat,
  title={IcySAT: Improved SAT-based attacks on cyclic locked circuits},
  author={Shamsi, Kaveh and Pan, David Z and Jin, Yier},
  booktitle={2019 IEEE/ACM International Conference on Computer-Aided Design (ICCAD)},
  pages={1--7},
  year={2019},
  organization={IEEE}
}

@inproceedings{tang2019openfpga,
  title={OpenFPGA: An opensource framework enabling rapid prototyping of customizable FPGAs},
  author={Tang, Xifan and Giacomin, Edouard and Alacchi, Aur{\'e}lien and Chauviere, Baudouin and Gaillardon, Pierre-Emmanuel},
  booktitle={2019 29th Int'l Conference on Field Programmable Logic and Applications (FPL)},
  pages={367--374},
  year={2019},
  organization={IEEE}
}

@inproceedings{kamali2023shell,
  title={SheLL: Shrinking eFPGA fabrics for logic locking},
  author={Kamali, Hadi M and Azar, Kimia Z and Farahmandi, Farimah and Tehranipoor, Mark},
  booktitle={2023 Design, Automation \& Test in Europe Conf. \& Exhibition (DATE)},
  pages={1--6},
  year={2023},
  organization={IEEE}
}

@article{baumgarten2010preventing,
  title={Preventing IC piracy using reconfigurable logic barriers},
  author={Baumgarten, Alex and Tyagi, Akhilesh and Zambreno, Joseph},
  journal={IEEE design \& Test of computers},
  volume={27},
  number={1},
  pages={66--75},
  year={2010},
  publisher={IEEE}
}

@inproceedings{kolhe2019custom,
  title={On custom LUT-based obfuscation},
  author={Kolhe, Gaurav and PD, Sai Manoj and Rafatirad, Setareh and Mahmoodi, Hamid and Sasan, Avesta and Homayoun, Houman},
  booktitle={Proceedings of the 2019 Great Lakes Symposium on VLSI},
  pages={477--482},
  year={2019}
}

@inproceedings{kamali2020interlock,
  title={InterLock: An intercorrelated logic and routing locking},
  author={Kamali, Hadi Mardani and Azar, Kimia Zamiri and Homayoun, Houman and Sasan, Avesta},
  booktitle={Proceedings of the 39th International Conference on Computer-Aided Design},
  pages={1--9},
  year={2020}
}

@inproceedings{kamali2019full,
  title={Full-lock: Hard distributions of sat instances for obfuscating circuits using fully configurable logic and routing blocks},
  author={Kamali, Hadi Mardani and Azar, Kimia Zamiri and Homayoun, Houman and Sasan, Avesta},
  booktitle={Proceedings of the 56th Annual Design Automation Conference 2019},
  pages={1--6},
  year={2019}
}

@inproceedings{guo2023evolute,
  title={Evolute: evaluation of look-up-table-based fine-grained ip redaction},
  author={Guo, Rui and Rahman, M Sazadur and Kamali, Hadi M and Rahman, Fahim and Farahmandi, Farimah and Tehranipoor, Mark},
  booktitle={2023 Design, Automation \& Test in Europe Conference \& Exhibition (DATE)},
  pages={1--6},
  year={2023},
  organization={IEEE}
}

@inproceedings{han2023functeller,
  title={$\{$FuncTeller$\}$: How Well Does $\{$eFPGA$\}$ Hide Functionality?},
  author={Han, Zhaokun and Shayan, Mohammed and Dixit, Aneesh and Shihab, Mustafa and Makris, Yiorgos and Rajendran, JV},
  booktitle={32nd USENIX Security Symposium},
  pages={5809--5826},
  year={2023}
}

@inproceedings{karmakar2024evaluating,
  title={Evaluating the robustness of large scale efpga-based hardware redaction},
  author={Karmakar, Praveen and Bharani, Marpina and Karfa, Chandan},
  booktitle={2024 37th International Conference on VLSI Design and 2024 23rd International Conference on Embedded Systems (VLSID)},
  pages={517--522},
  year={2024},
  organization={IEEE}
}

@inproceedings{sathe2023mantis,
  title={Mantis: Machine learning-based approximate modeling of redacted integrated circuits},
  author={Sathe, Chaitali G and Makris, Yiorgos and Schafer, Benjamin Carrion},
  booktitle={2023 Design, Automation \& Test in Europe Conference \& Exhibition (DATE)},
  pages={1--6},
  year={2023},
  organization={IEEE}
}

@inproceedings{rezaei2022evaluating,
  title={Evaluating the security of eFPGA-based redaction algorithms},
  author={Rezaei, Amin and Afsharmazayejani, Raheel and Maynard, Jordan},
  booktitle={Proceedings of the 41st IEEE/ACM Int'l Conf. on Computer-Aided Design},
  pages={1--7},
  year={2022}
}

@inproceedings{shamsi2018cross,
  title={Cross-lock: Dense layout-level interconnect locking using cross-bar architectures},
  author={Shamsi, Kaveh and Li, Meng and Pan, David Z and Jin, Yier},
  booktitle={Proceedings of the 2018 Great Lakes Symposium on VLSI},
  pages={147--152},
  year={2018}
}

@article{luu2014vtr,
  title={VTR 7.0: Next generation architecture and CAD system for FPGAs},
  author={Luu, Jason and Goeders, Jeffrey and Wainberg, Michael and Somerville, Andrew and Yu, Thien and Nasartschuk, Konstantin and Nasr, Miad and Wang, Sen and Liu, Tim and Ahmed, Nooruddin and others},
  journal={ACM Transactions on Reconfigurable Technology and Systems (TRETS)},
  volume={7},
  number={2},
  pages={1--30},
  year={2014},
  publisher={ACM New York, NY, USA}
}

@inproceedings{mcmurchie1995pathfinder,
  title={PathFinder: A negotiation-based performance-driven router for FPGAs},
  author={McMurchie, Larry and Ebeling, Carl},
  booktitle={Proceedings of the 1995 ACM third international symposium on Field-programmable gate arrays},
  pages={111--117},
  year={1995}
}

@inproceedings{sweeney2020modeling,
  title={Modeling techniques for logic locking},
  author={Sweeney, Joseph and Heule, Marijn JH and Pileggi, Lawrence},
  booktitle={Proceedings of the 39th International Conference on Computer-Aided Design},
  pages={1--9},
  year={2020}
}

@inproceedings{alrahis2021untangle,
  title={UNTANGLE: Unlocking routing and logic obfuscation using graph neural networks-based link prediction},
  author={Alrahis, Lilas and Patnaik, Satwik and Hanif, Muhammad Abdullah and Shafique, Muhammad and Sinanoglu, Ozgur},
  booktitle={2021 IEEE/ACM Int'l Conference On Computer Aided Design (ICCAD)},
  pages={1--9},
  year={2021},
  organization={IEEE}
}

@inproceedings{shamsi2019kc2,
  title={KC2: Key-condition crunching for fast sequential circuit deobfuscation},
  author={Shamsi, Kaveh and Li, Meng and Pan, David Z and Jin, Yier},
  booktitle={2019 Design, Automation \& Test in Europe Conference \& Exhibition (DATE)},
  pages={534--539},
  year={2019},
  organization={IEEE}
}

@article{das2025soar,
  title={SOAR: Secure Once, Adapt at Runtime with eFPGA-Based Redaction for IP Protection},
  author={Das, Voktho and Azar, Kimia and Kamali, Hadi},
  journal={IEEE Access},
  year={2025},
  publisher={IEEE}
}

\end{document}